\documentclass[linenumbers]{aastex631}

\begin{document}

\nolinenumbers
\title{ M-$\sigma$ RELATIONS ACROSS SPACE AND TIME}

\author{David Garofalo}
\affiliation{Department of Physics, Kennesaw State University \\}

\author{Damian J. Christian}
\affiliation{Department of Physics and Astronomy, California State University, Northridge \\}

\author{Chase Hames}
\affiliation{Department of Physics, Kennesaw State University \\}

\author{Max North}
\affiliation{Department of Information Systems, Kennesaw State University}
\author{Keegan Thottam$^{1}$}

\author{Samuel Nazaroff$^{1}$}

\author{Alisaie Eckelbarger$^{1}$}

\begin{abstract}

Feedback from active galactic nuclei (AGN) has long been invoked to explain the correlation between black hole mass and stellar velocity dispersion (M-$\sigma$) discovered in low redshift galaxies. We describe the time evolution of AGN in the M-$\sigma$ plane based on our gap model (Garofalo, Evans \& Sambruna 2010) for black hole accretion and jet formation illustrating a fundamental difference between jetted and non-jetted AGN. While the latter tend to evolve diagonally upward with black hole mass increasing along with stellar dispersion, we show that jetted AGN tend on average to move initially more upwards because their effect on velocity dispersion is weaker than for non-jetted AGN. But this initial phase is followed by a shift in the nature of the feedback, from positive to negative, a transition that is more dramatic on average in denser cluster environments. The feedback gets its kick from tilted jets which shut down star formation but increase velocity dispersion values. As this change in the nature of the feedback takes tens of millions to hundreds of millions of years, these cluster, merger-triggered jetted AGN, will evolve more upwards for up to order $10^{8}$ years, followed by an extremely long phase in which low excitation progressively slows black hole growth but dramatically affects stellar dispersion. As a result, powerful jetted AGN evolve for most of their lives almost horizontally on the M-$\sigma$ plane. The prediction is that strongest AGN feedback on stellar dispersion is a late universe phenomenon with M87 a good example. We show how jetted and non-jetted AGN parallel the Sersic and core-Sersic galaxy paths in the M-$\sigma$ plane found by Sahu et al (2019) and to a prediction that jetted quasars are not core-Sersic galaxies as found for lower redshift jetted AGN.

\end{abstract}


\section{Introduction} 

 For a quarter century, evidence has been building that massive black holes occupy the centers of galaxies and that stellar velocity dispersions in the bulge are tightly connected with black hole mass via the so-called M-$\sigma$ relation (Magorrian et al 1998; Gebhardt et al 2000; Ferrarese \& Merritt 2000; see Kormendy \& Ho 2013 for a review)

\begin{figure} [ht]
   \includegraphics[scale=1]{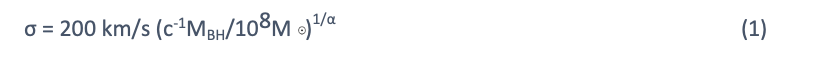}
   
\end{figure}


with c and $\alpha$ constants with values 3 and 4, respectively (e.g. Zubovas \& King 2019).  We plot equation (1) in Figure 1. The nature of the correlation has been the subject of debate but no clear picture has emerged. The dominant idea is that AGN feedback is responsible (e.g. Silk \& Rees 1998). In Figure 1 we also plot the galaxies of Gultekin et al 2009.  Because the majority of AGN do not produce powerful jets, it is not unreasonable to assume that the majority of these objects were once radio quiet AGN.  As a result, the fit and these objects give us a sense of how the non-jetted AGN populate the M-$\sigma$ diagram.

We apply the time evolution in the gap paradigm for black holes (Garofalo, Evans \& Sambruna 2010), to understand the nature of the M-$\sigma$ relation for different types of AGN. We show that whereas non-jetted and generally more core-Sersic AGN move both upwards and to the right in the M-$\sigma$ plane, jetted and generally more Sersic AGN \footnote[1]{Sérsic and core-Sérsic galaxies are defined by two distinct relations, $M_{BH}\propto\sigma^{5.75+-0.34}$ and  $M_{BH}\propto \sigma^{8.64+-1.10}$, respectively (Sahu et al. 2019).}, tend to move more upwards initially, followed by mostly horizontal motion, which on average is environment-dependent. We identify characteristic timescales, characteristic black hole mass growth, and characteristic velocity dispersion, in order to explain the physical origin of this difference as a key component of the radio loud/radio quiet dichotomy. This current work focuses on a more qualitative picture of these parameters as we build on our models for future more quantitative descriptions. In Section 2 we address the time dependent feedback on star formation, accretion, and stellar velocities of jetted and non-jetted AGN and compare to observations. In Section 3 we conclude.

\section{Paths of AGN in the M-$\sigma$ plane}

Two kinds of black hole feedback contribute to enhancing the stellar dispersion in the gap paradigm (Garofalo, Evans \& Sambruna 2010): 1) radiatively-driven accretion disk winds and 2) tilted jets, with the second dominant but present only in a subclass of AGN. In this section, we argue that these two elements drive two different M-$\sigma$ relations and make predictions for observations. The model describes the origin of both FRI and FRII jets, with the former less powerful on average that often experience entrainment due to the surrounding medium while the latter are on average more powerful and collimated and visible in hotspots (Fanaroff $\&$ Riley 1974). While FRII jets enhance the star formation rate and may alter the state of accretion, they do not contribute to stellar dispersion. This is due to the fact that they are not tilted jets. Because more massive black holes produce greater AGN feedback, there is a dependence on black hole mass that translates into a dependence on environment that we now describe. 

 In Figure 2 we consider a high spinning black hole that is triggered into an AGN either by secular processes in a spiral galaxy, or by a merger. The cold gas is funneled into the nucleus and eventually forms a thin, radiatively efficient disk, that co-rotates with the black hole in the sense that the angular momentum vectors of the black hole and disk are aligned. This is the most likely configuration that results from the Bardeen-Petterson effect (Bardeen \& Petterson 1974). Because the innermost stable circular orbit (ISCO) is closer to the black hole than when the angular momenta are anti-aligned, the radiative efficiency of the accretion disk is high and this leads to powerful, radiatively-driven winds, as shown by the arrows in red in Figure 2. Accretion onto the black hole further spins the black hole up and the ISCO, therefore, moves further inward toward the black hole. The value of black hole spin is shown on the right and the rate of star formation is shown on the left. The feedback on the star formation rate was described in Singh et al 2021. Such systems are subject to jet suppression (Ponti et al 2012; Garofalo, Evans \& Sambruna 2010; Nielsen \& Lee 2009), which kicks in around a spin of 0.7, and therefore become radio quiet or jetless AGN. Because the spin is above 0.7 in Figure 2, such objects are born jetless. Our model has a key parameter of spin alignment and not just the general agreement that radio loud (jetted) AGN tend to host SMBH exceeding $10^{8}$ \(M_\odot\) (Chiaberge \& Marconi 2011). Accretion adds to the mass of the black hole while the radiative wind contributes to the velocity dispersion (e.g. Silk \& Rees 1998). The cartoons in the column labeled $\sigma$ capture the relative scale of the effect of the wind on the stellar dispersion via the length of the white arrows. The arrows in Figure 2 are of intermediate length compared to those in Figures 3, 4, and 5. Co-rotating systems, again, constitute the majority of AGN, and Figure 2 objects purport to explain equation 1 and are captured as the blue line in both Figures 1 and 6. It is important to emphasize that Figures 2, 3, 4, and 5 are meant to produce a holistic (but not quantitative) picture of differences in black hole feedback. This does not preclude quantitative estimates for feedback and timescales for that feedback. While Figure 1 has low redshift observational data, Figure 6 describes the evolution in time on the M-$\sigma$ plane for two black holes.  The slope of the blue line captures the increase in black hole mass that is associated with an increase in the stellar dispersion, through the disk wind, in such systems.  If the black hole spin is lower than in Figure 2, an FRI jet may exist. This is where the model addresses so-called jetted narrow line Seyfert 1 but does not add anything deep to the ideas here because jet feedback is not great. In other words, black holes that are triggered in corotation for any spin value tend to grow black holes and affect stellar dispersion in a mild way, unlike for black holes that are triggered in counter-rotation - and especially for those in richer environments - as we now describe.
	
In Figure 3 we show the average evolution in time of black holes in isolated environments such that their black holes are not the most massive but are triggered by mergers, start with the same spin value as in Figure 2, but differ in the orientation of the accretion disk, which ends up in a counterrotating configuration with respect to the black hole. I.e. the angular momenta of black hole and disk are anti-aligned. These are difficult to form (which is why Figure 2 captures most post-merger systems in these isolated field environments as well) because of stability issues (Garofalo, North, Belga, Waddell 2020; King et al 2005) and are therefore a minority. Counterrotation is key to the production of FRII jets (Garofalo, Evans \& Sambruna 2010), which results from maximizing both the Blandford-Znajek (BZ) and Blandford-Payne (BP) effects.  Notice that such systems are not hosted in spiral galaxies because they are the product of mergers.  FRII jets increase the star formation rate as shown in the left column (e.g. Kalfountzou et al 2014; Singh et al 2021). Although the radiative wind from the disk contributes to the stellar dispersion, it is less effective than in Figure 2 systems because the radiative efficiency of the disk is lower in counterrotation due to the larger value of the ISCO compared to corotation (compare lowest panels of Figure 2 and Figure 3). Notice how the inner edge of the disk evolves from large value (lowest panel of Figure 3) to low value (highest panel of Figure 3).  

     As the black hole spin changes in time due to accretion, the radiative wind increases in tandem with the decreasing distance of the ISCO. But when the system enters the corotating phase, notice the existence of an FRI jet and an associated increase in the size of the white arrows for the column labeled $\sigma$. FRI jets, in fact, have the most dominant effect on stellar dispersion. This is due to the fact that in the transition through zero black hole spin, the disk experiences a shift in orientation (Garofalo, Joshi et al 2020). Once the black hole spin increases sufficiently (a = 0.2) such that an FRI jet is produced, it is also likely to be tilted with respect to the previous FRII jet, and the interstellar medium is now sprayed head on by the jet. This is the reason why FRI jets suppress star formation (the Roy Conjecture – Garofalo, Moravec et al 2022) while having a positive feedback effect on stellar dispersion. 
     
     The possibility of a tilted jet forming in the transition from counter-rotation to co-rotation is key to understanding the different behavior in the M-$\sigma$ plane for jetted AGN. But these isolated systems described in Figure 3 remain radiatively efficient and are therefore subject to the same jet suppression at high spin of Figure 2, hence the jet disappears. Once the FRI jet disappears (highest panel of Figure 3), the dominant effect on stellar dispersion drops and the white arrows in the $\sigma$ column shorten again. The period during which the FRI jet exists is a few tens of millions of years. Compared to Figure 2, Figure 3 systems appear to experience lower contributions to stellar dispersion earlier, and then make up for that later with the FRI jet. Because of the larger gap region between disk and black hole in Figure 3 systems, the disk wind is weaker compared to Figure 2 systems which means that Figure 3 AGN move more upwards in the M-$\sigma$ plane initially compared to Figure 2 AGN (This is also true of the more dominant jetted systems we will describe next and is the reason why the red curve (RLQ) in Figure 6 is initially leftward of the blue curve (RQQ)). The counterrotating black holes of Figure 3, therefore, must possess more massive black holes to achieve a given $\sigma$ value. Because of the dominant effect of FRI jets, such systems will eventually shift to more horizontal motion in the M-$\sigma$ diagram compared to Figure 2 systems once corotation is achieved. We now move on to systems whose FRI jets are more dominant.

     In Figure 4, we show the average time evolution of black hole systems that formed via mergers in denser environments that lead to initial conditions that involve more massive black holes compared to Figure 3 but that also involves counterrotation (here too, most post-merger systems end up in corotation as in Figure 2). The same initial conditions of Figure 3 explains the FRII jet and the high rate of star formation as well as the weaker contribution to stellar dispersion compared to Figure 2. Crucially, however, the FRII jet feedback is greater than that of Figure 3 and this leads to the change in accretion state at late times (see advection dominated accretion flow (ADAF) in the upper panel – Garofalo, Evans \& Sambruna 2010). Because of the absence of a thin disk at late times, jet suppression fails and the tilted FRI jet lives on as long as accreting fuel persists. This enables such systems to further enhance their stellar dispersions as seen by the large white arrows in the $\sigma$ column in the two upper panels of Figure 4. It is important to note that ADAF accretion feeds black holes at lower rates compared to thin disks (lower than $10^{-2}$ that of the Eddington rate). Therefore, upward migration slows down while horizontal motion speeds up on the M-$\sigma$ diagram. Figure 4 systems evolve in a similar way to Figure 3 systems on the M-$\sigma$ plane except at late times when they can affect their stellar dispersions in a more dominant way due to the tilted FRI jet. 

     In Figure 5, we show the average time evolution of the most massive black hole systems triggered in the densest environments into counterrotation (Again, the more likely post-merger state is accretion in corotation as in Figure 2). The reader can anticipate that feedback will be similar to Figures 3 and 4, only more dramatic due to the triggering of the most massive black holes. Because of the dominant effect of FRII jets, such systems evolve rapidly away from thin disks and into ADAF states, which means that once the spin is large enough for the FRI jet to form, it will be part of the longest of any AGN subclass (e.g. M87). Despite a period in which stellar dispersion is not affected (because the disk is an ADAF and not radiatively efficient), the contribution to stellar dispersion from FRI jets is most dominant and is captured by the red path in Figure 6. Such systems move for billions of years mostly horizontally in the M-$\sigma$ plane. 
     
     In order to capture the basic differences between the two kinds of M-$\sigma$ relations that are implied by Figures 2-5, we show, in Figure 6, characteristic evolution timescales for the objects represented by Figure 2 (blue path) and Figures 4 and 5 (red path). In the lowest panels of Figures 2 and Figures 4 and 5, we are dealing with Eddington-limited accretion onto rapidly rotating black holes in corotation and counterrotation. For purposes of direct comparison, we pick a black hole that is triggered with the same mass ($10^{9}$ solar masses). Because the inner edge of the disk is closer to the black hole represented in Figure 2, its disk wind is more powerful than that of Figures 4 and 5, which manifests itself with a larger $\sigma$ value (the blue curve is slightly rightward of the red one). Counterrotation lasts $10^{7}$ years for Figure 4 objects and 4 x $10^{8}$  years for Figure 5 objects because the accretion rates are at the Eddington limit for Figure 4 objects and at $10^{-2}$ the Eddington limit within 4 million years for Figure 5 objects. Given this information, one can evaluate the total accreted mass during this phase. In the transition through zero spin, Figure 4 and 5 systems experience absence of the Bardeen-Petterson effect and a possible tilt in the accretion disk, which due to spin-up then leads to a tilted FRI jet and ADAF accretion.  This generates a dramatic shift in the M-$\sigma$ plane, with slower black hole growth that begins around $10^{-3}$ the Eddington rate and drops two or more orders of magnitude below that over order a billion years, but larger velocity dispersions as shown in Figure 6 with the red curve tilting rightward. These are the types of paths that lead to the low redshift M-$\sigma$ relation. In short, our model teases out two different behaviors associated with jetted and non-jetted AGN.

 \begin{figure}
   \includegraphics[scale=0.5]{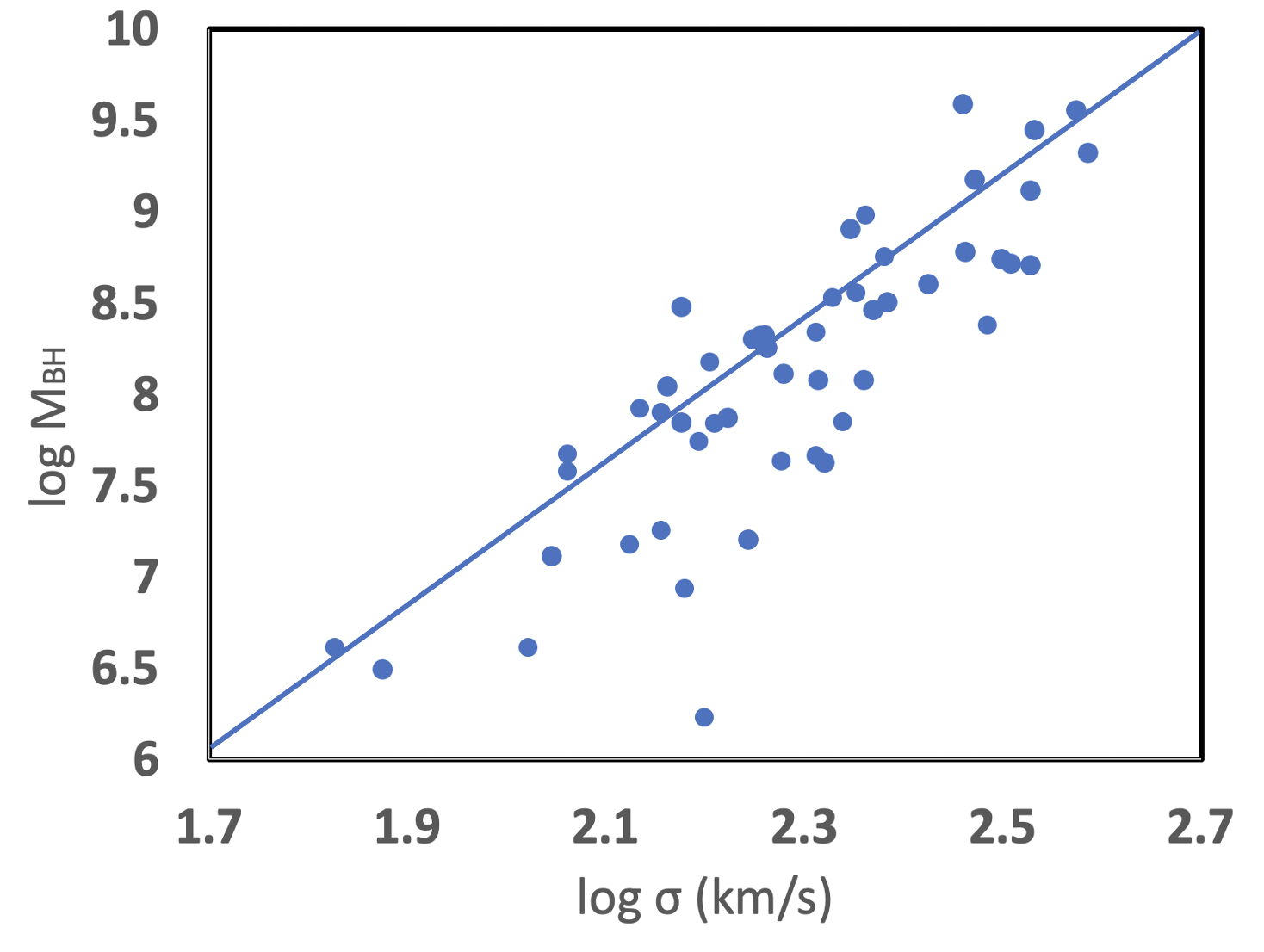}
   \caption{The M-$\sigma$ relation of Gultekin et al (2009). Equation 1 is over-plotted as the blue-line. $M_{BH}$ is the mass of the black hole divided by 1 solar mass.}
\end{figure}

\begin{figure}[t]
        \includegraphics[scale=0.4]{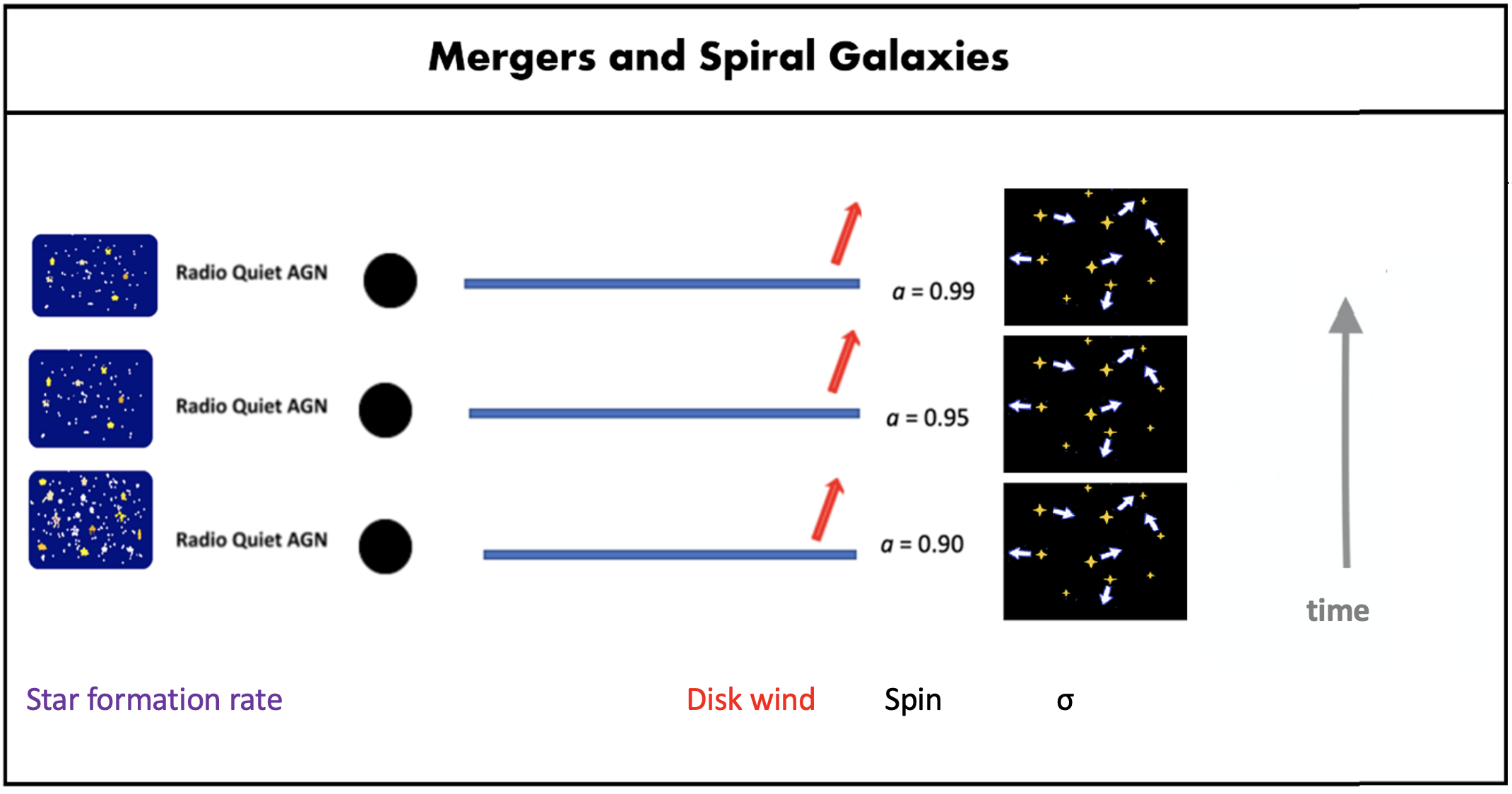}
        \caption{Most AGN will be triggered into corotation between black hole and accretion disk, whether it results from a merger or secular processes. If such systems are rapidly spinning black holes, they will simply spin the black hole up further. The radiative wind from the disk will increase slightly over time and its effect on the stellar dispersion will increase. }
\end{figure}

   \begin{figure} 
    \includegraphics[scale=0.7]{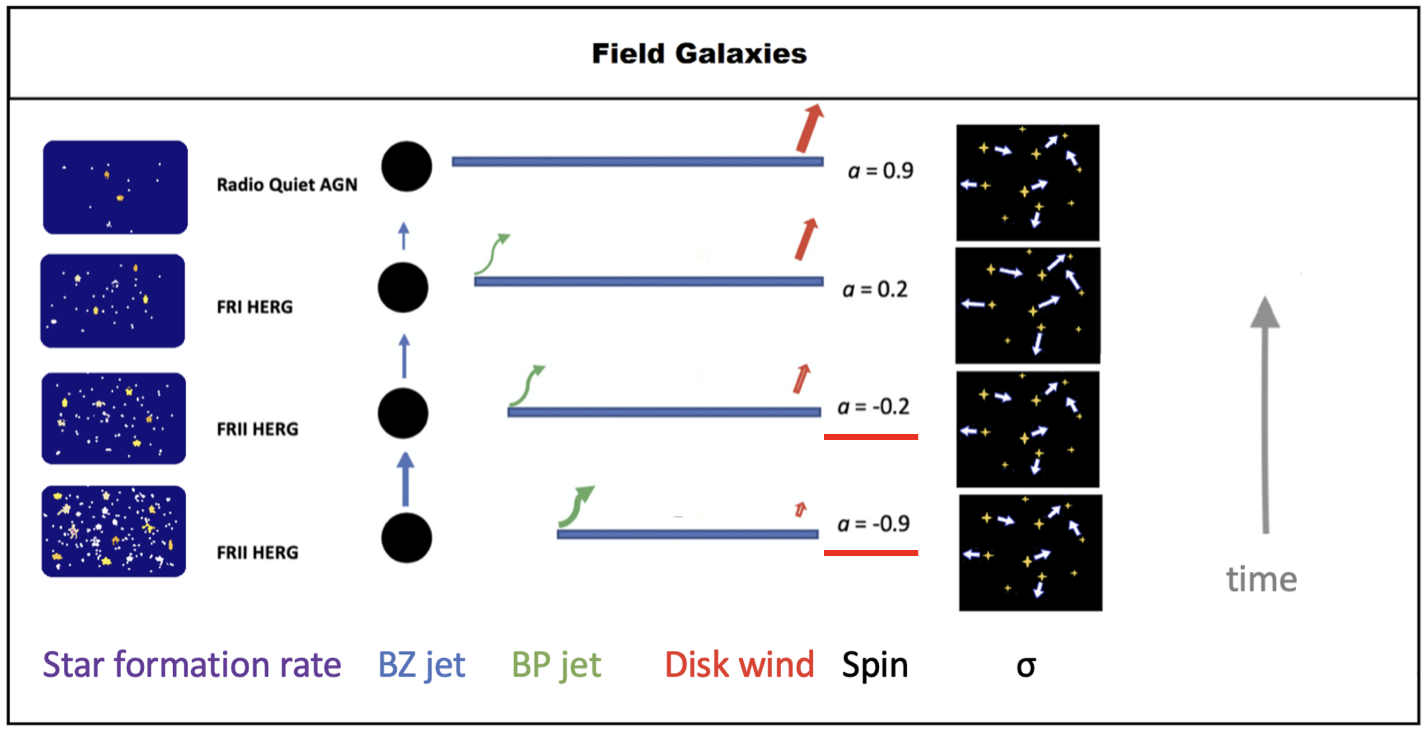}
     \caption{For the subclass of AGN that are in isolated environments such that the merger leads to counter-rotation, the radiative wind from the disk will contribute relatively less to the stellar velocity dispersion compared to Figure 2 because counter-rotating disks have lower efficiency and thus weaker disk winds. A short-lived FRI jet will also increase $\sigma$ for a few tens of millions of years until jet suppression kicks in, and the system behaves like the high-spinning black holes of Figure 2 thereafter. HERG = high excitation radio galaxy.  Counter-rotating phase underlined in red.}
    \end{figure}

   \begin{figure}[h]
        \includegraphics[scale=0.6]{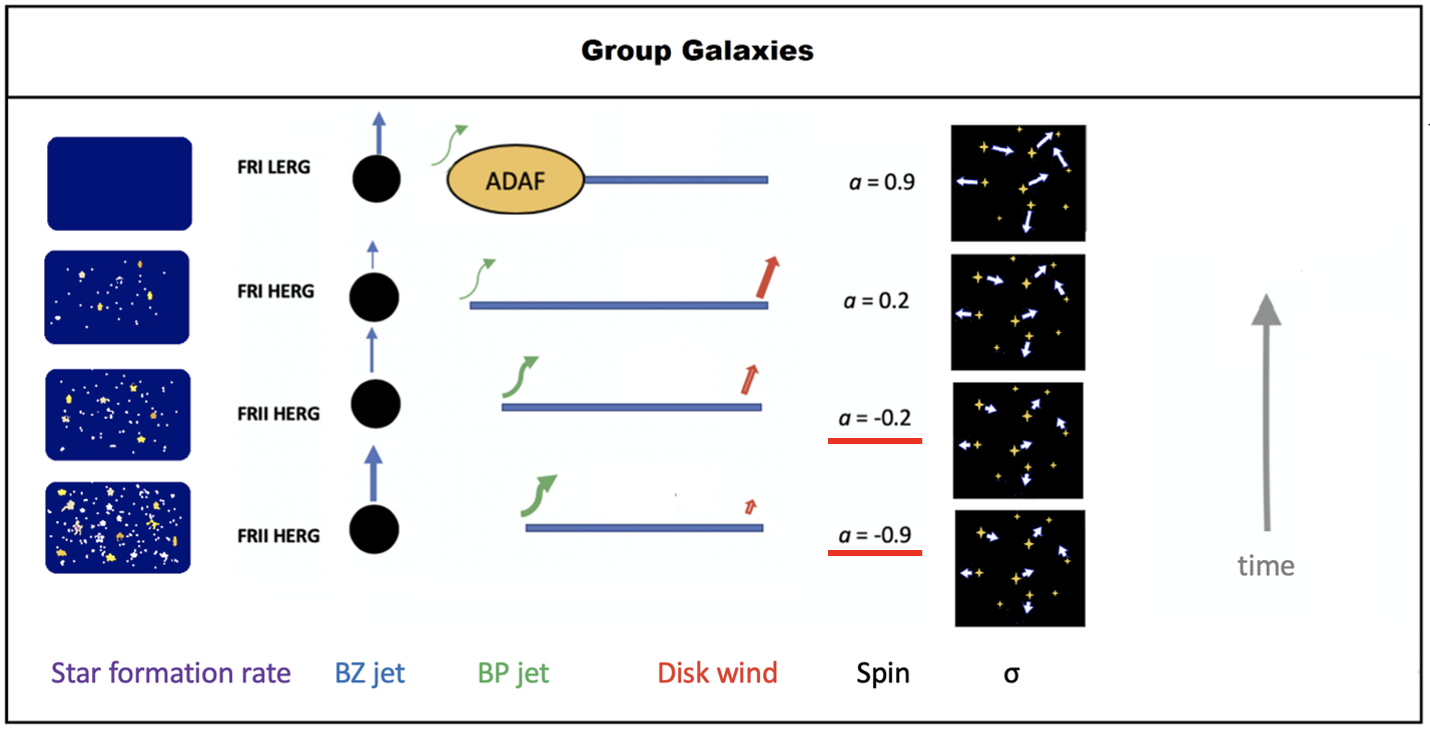}
        \caption{More massive black holes on average in denser group environments results in greater AGN feedback on the disk which evolves into an ADAF. Such systems differ from Figure 3 in that the FRI jet persists which leads to a prolonged effect on the stellar dispersion and increased horizontal red paths in Figure 6. LERG = low excitation radio galaxy. Counter-rotating phase underlined in red.}
    \end{figure} 

    \begin{figure*}[h]
        \includegraphics[scale=0.8]{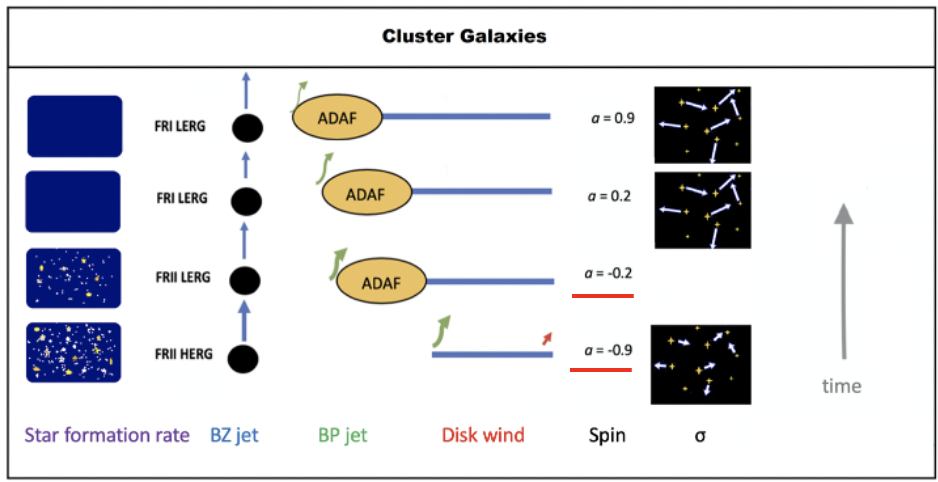}
        \caption{Strongest AGN feedback and rapid evolution in ADAF accretion states which allows the FRI jet to emerge early and have a dominant effect on stellar dispersion. The feedback on stellar dispersion is weak in the second panel from bottom, hence no image is shown. Such systems move more horizontally in Figure 6 than any other AGN subclass.  }
    \end{figure*}

\begin{figure*}
        \includegraphics[scale=0.5]{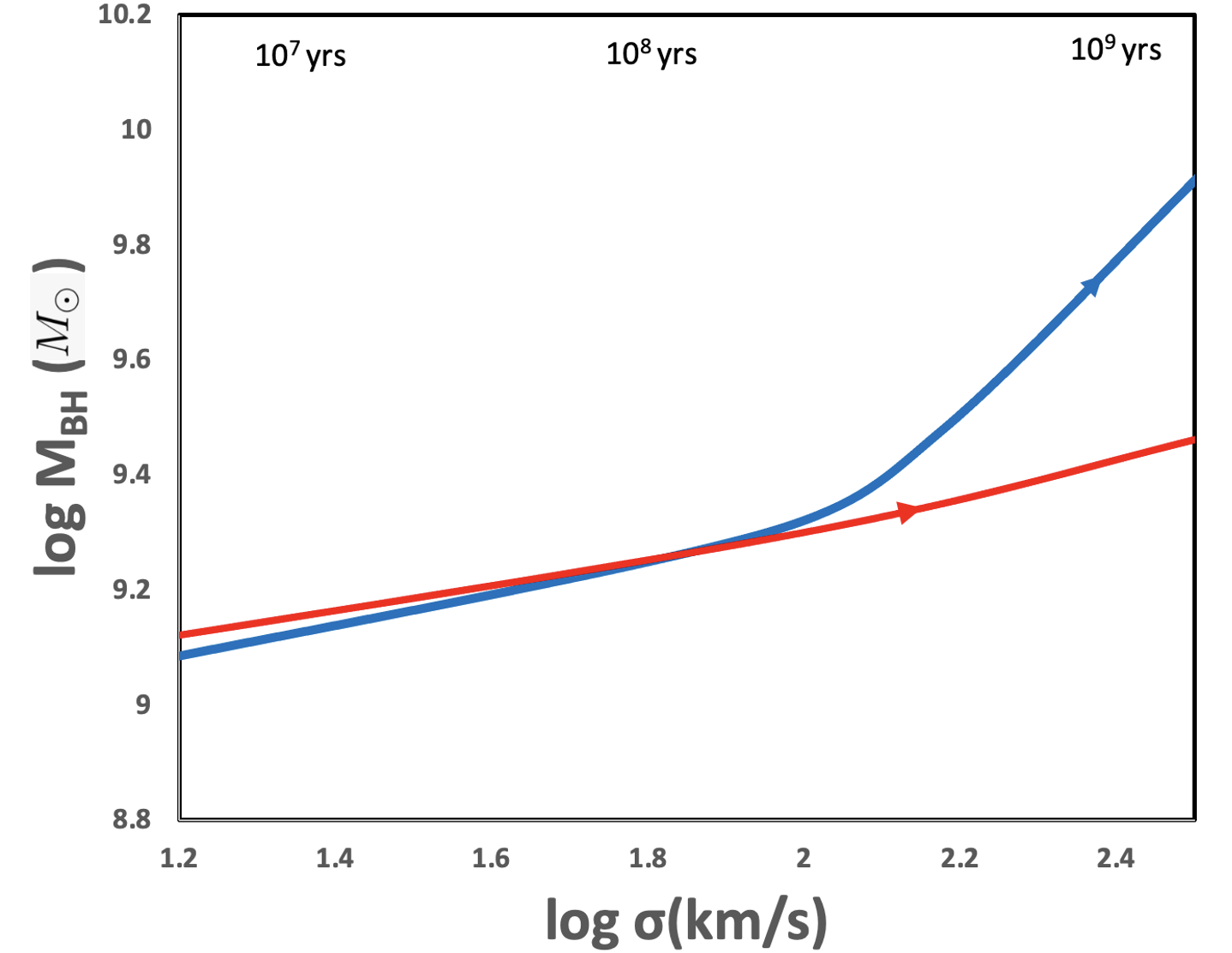}
        \caption{Theoretical M-$\sigma$ paths. Blue line: Radio quiet quasars (RQQ). Red path: Radio loud quasars (RLQ).  Both systems begin with a rapidly spinning, billion solar mass black hole, surrounded by a thin accretion disk in corotation (blue) and counterrotation (red). The blue system corresponds to Figure 2 while the red captures average evolution between Figures 4 and 5. Because the ISCO is smaller for the blue system, the disk wind more strongly affects stellar dispersion. Feedback on stellar dispersion is reversed when the tilted FRI jet is generated in the red systems. The change in the black hole engine depends on the value of the spin which changes due to accretion in a quantifiable way. Characteristic timescales for those changes are labeled on the upper horizontal axis. }
    \end{figure*}

\begin{figure*}
        \includegraphics[scale=0.5]{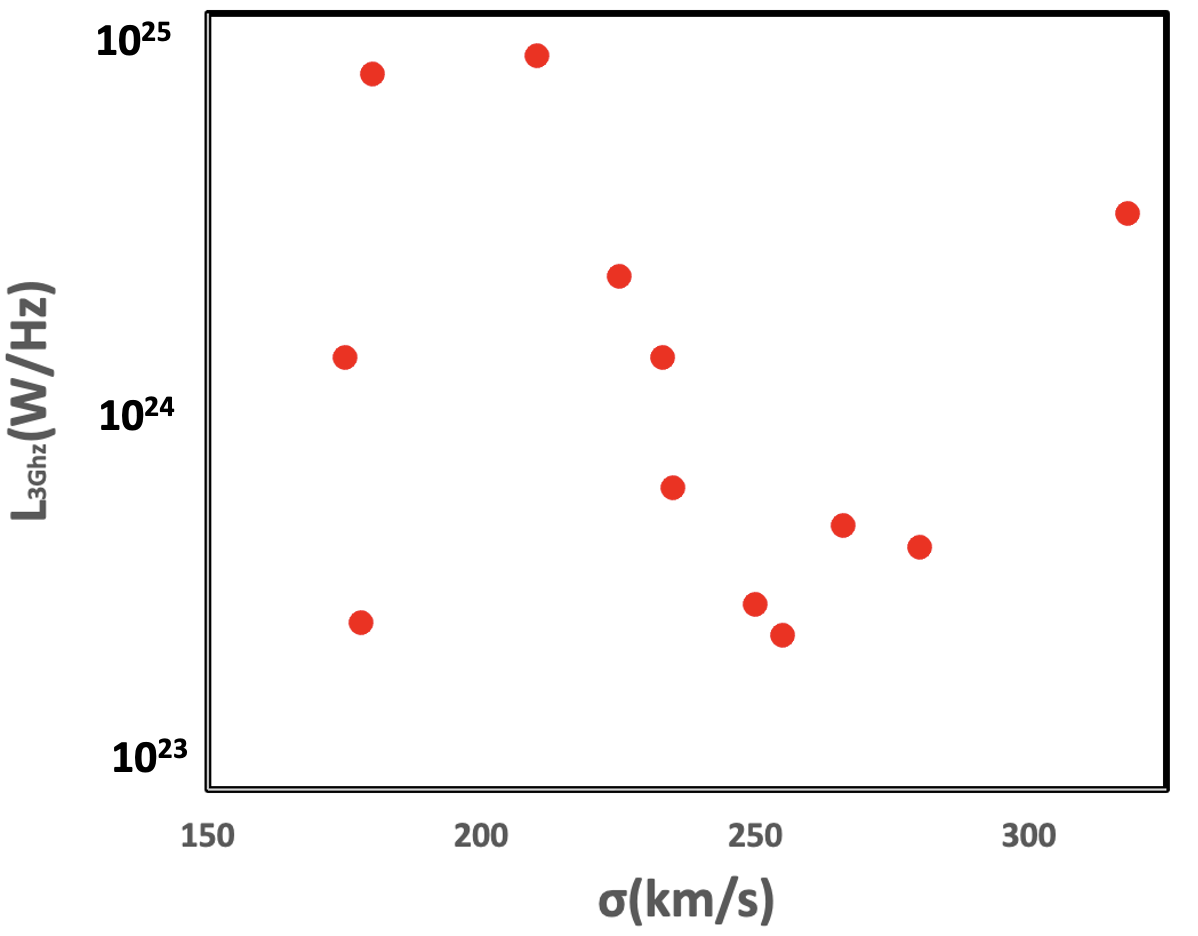}
        \caption{3 GHz radio luminosity as a function of stellar dispersion from Barisic et al (2017) showing absence of correlation between FRII jet power and stellar velocities. }
    \end{figure*}

    \begin{figure*}
        \includegraphics[scale=0.9]{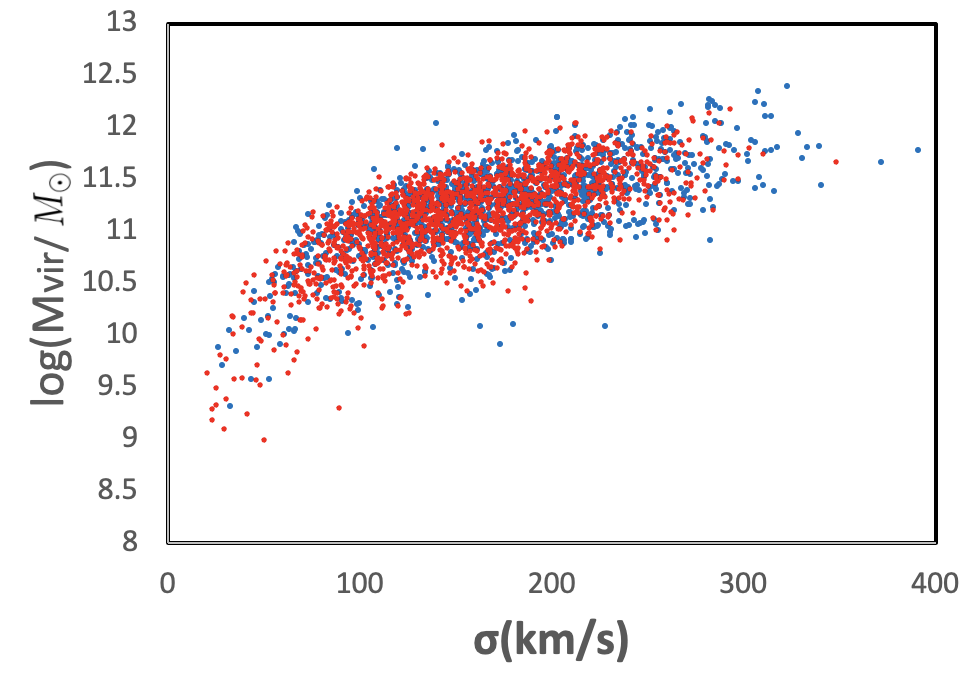}
        \caption{Virial mass versus stellar dispersion from van der Wel et al (2021) as evidence for the behavior predicted in Figure 6. Objects with redshift between 0.6 and 0.8 in red while objects with redshift between 0.8 and 1 in blue.}
    \end{figure*}
    
         The model features described above predict that  for large(r) data sets, FRI LERG will on average tend to have larger $\sigma$ values than FRII HERG. FRII HERG are predicted to be phases in which the accretion rates are high and the black hole mass is thus increasing at high rates compared to latter stages. I.e. they are represented in the initial part of the red curve in Figure 6 with largest slope. Hence, on average, FRII HERG should also have lower black hole masses than FRI LERG (e.g. Miraghaei \& Best 2017).

         Let us now describe the effects of AGN feedback on star formation as depicted in the first column of Figures 2-5. Along with more rightward location on the M-$\sigma$ diagram, FRI LERG systems should also experience the lowest star formation rates of any other subclass of AGN (Figure 5 left column of upper two panels). FRII HERG systems, on the other hand should experience high star formation rates (left column and lowest panels of Figures 3, 4, and 5). This has indeed been shown to be the case in the star formation – stellar mass plane including the intermediate FRII LERG systems (Garofalo \& Pouliasis 2022). Because FRII jets do not enhance stellar dispersion while FRI jets do, jet power as a function of stellar dispersion should not show any correlation if FRII HERG and FRII LERG are included.  We choose only the FRII candidates from Figure 6 of Barisic et al 2017 where they plot stellar velocity dispersion and jet powers. For FRII objects and in Figure 7, we show the radio luminosity (a proxy for jet power) versus stellar dispersion, supporting an absence of correlation between jet power and stellar dispersion. Because stellar dispersion increases over time once the FRI jet is triggered, the prediction is that a correlation should instead exist between the dynamical FRI jet lifetime and velocity dispersion, unlike for FRII sources.

         The FRII phases in our figures are relatively short in the less dense environments and associated with rapid black hole growth. Because the accretion rate drops orders of magnitude below the Eddington limit during FRI ADAF phases (upper panels of Figures 4 and 5), black hole growth slows down dramatically within 10-100 million years from the triggering of the radio galaxy. If we could capture the evolution with rapid snapshots in the early phase, we would expect to see an early large slope followed by a bend toward a lower slope to capture the dominant effect of tilted FRI jets on stellar dispersion at late times . But this requires exploring M-$\sigma$ at higher redshift. 
         The bending in the M-$\sigma$ plane experienced by radio loud quasars as they evolve into low excitation systems shown in Figure 6 is difficult to tease out of an observational M-$\sigma$ diagram. It should be possible to capture this effect with more data and higher redshift. This has been done in van der Wel et al. (2021), although with virial mass. We take their virial masses and stellar dispersions from LEGA-C with galaxies at redshifts $0.6 < z < 1$ and plot them in Figure 8. We see the bending described in Figure 6. While caution is in order here due to the fact that virial masses are often systematically larger than dynamical masses, this should not affect the bend.

         {In our final analysis, we note that we are not the first to suggest different M-$\sigma$ relations. That separate M-$\sigma$ relations for Sersic and core-Sersic galaxies exist was discovered recently (Sahu, Graham $\&$ Davis 2019). They make a fit for both populations in the M-$\sigma$ plane and report them in equations 4 and 6 which we plot in Figure 9. Notice that core-Sersic galaxies tend to behave in a way that is similar to the RQQ in Figure 6.  Core-Sersic galaxies tend to have dimmer cores compared to Sersic galaxies (Graham et al 2003), the origin of which is thought to be that such systems are formed as the result of a dry merger with less gas in the central region and therefore the merging black holes in the nucleus require stars to solve the last parsec problem (Milosavjlevic $\&$ Merritt 2001). It has long been recognized that the last parsec problem is less problematic for counterrotating disk configurations (Nixon et al 2011). We are on a spectrum here, where both wet and dry mergers that form co-rotating disk configurations in their nuclei (again, this is the majority of post-merger configurations for the nuclear disk), while struggling to merge their black holes, will appeal less to stars in wet mergers than in dry ones due to the availability of cold gas in the former. Counterrotating configurations, on the other hand, will not need as much help to coalesce the black hole binary. This means that all mergers that lead to co-rotating disk configurations will struggle to merge their black holes and will have to eject from the nucleus a mass in stars that can be up to the order of the binary mass (Milosavljevic $\&$ Merrit 2001), significantly reducing the density in the core. As a result, co-rotating accretion disks around newly merged black holes will both have lower core density and thus brightness (core-Sersic profiles) and will tend to be non-jetted AGN. RQQ are the brightest of non-jetted AGN because of the abundant amount of cold gas that fuels the AGN, and therefore will have had an opportunity to coalesce their binary black holes using gas, dust, and stars, such that their core-Sersic profiles should be different from similar coalescing black hole binaries that formed in dry mergers in the sense that less stars are evacuated and the core-Sersic profile less steep. Note that Figure 6 deals with extremes, i.e. most radio loud (or jetted) and most radio quiet (or non jetted). Thus, such extremes are not expected to fully explain Figure 9. It is interesting to note that the red curve of Figure 6 (RLQ) would bend less away from the blue curve (RQQ) if we incorporated less extreme objects. Along those same lines, note how Figure 9 curves intersect at lower black hole mass compared to Figure 6. If we think of extreme jetted AGN or RLQ, such objects are growing their black holes at the most rapid rate while affecting their dispersions in the weakest way so it makes sense that overtaking RQQ will occur at higher mass as seen in Figure 6.} 

         While extreme RQQ and RLQ are prescribed to have larger and smaller M-$\sigma$ slopes, respectively, less extreme AGN will distribute themselves in-between. Since the majority of post-merger black holes end up in co-rotating disk configurations, if we consider systems such as those in Figure 2 but with nascent black hole spin values in the range 0.2 $<$ \emph a $<$ 0.7, there is an absence of jet suppression and a jetted AGN will form. Given the initial co-rotating disk configuration, such systems are expected to be core-Sersic galaxies and therefore the division between Sersic profile and radio loudness begins to break down. We claim that such jetted AGN are the ones explored in Capetti \& Brienza (2023) and that a mixing, therefore, exists among jetted AGN such that some have evacuated cores and others do not. The important point to notice is that the jetted AGN subclass with evacuated cores cannot be the extreme RLQ that we plotted in Figure 6. The absence of counter-rotation in this core-Sersic class of jetted AGN ensures absence of low excitation and the absence of tilted jets, and therefore absence of strong $\sigma$ values, features that increase the slope of such objects on the M-$\sigma$ plane. In other words, as we consider less extreme jetted AGN, they will distribute themselves on the M-$\sigma$ plane with slopes that are in-between the extreme RLQ and RQQ and we will expect a mixing of core-Sersic and Sersic profiles with jetted and non-jetted AGN. In addition, they will also necessarily not experience the relatively low star formation rates associated with tilted jet suppression of star formation. These are the objects described recently in Capetti \& Brienza (2023) but see also Hamilton (2010).

         \begin{figure*}
        \includegraphics[scale=0.9]{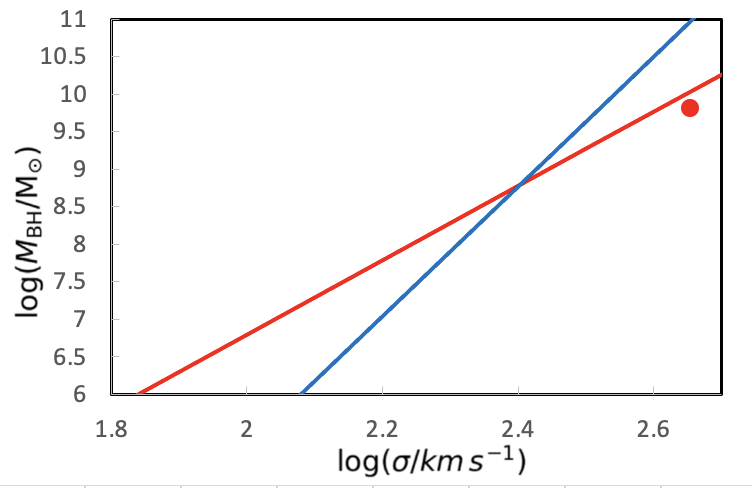}
        \caption{Fits for the M-$\sigma$ relation for Sersic (red) and core-Sersic (blue) ellipticals (equations 4 $\&$  6) from Sahu, Graham \& Davis 2019. The red dot represents M87, an FRI LERG in a cluster environment.}
    \end{figure*}


\section{Conclusions}
     The strong correlation of central black hole mass and stellar velocity dispersion (M-$\sigma$ plane) has been the subject of debate, but no clear picture has emerged. The gap paradigm for black holes has given us a roadmap for understanding the radio loud/radio quiet dichotomy, with the former the result of original counterrotation between black hole and accretion disk (Garofalo, Evans \& Sambruna 2010). We have shown here how to understand the location of different subclasses of AGN in the M-$\sigma$ plane based on a shift from relatively stronger feedback in the jetted subclass compared to non-jetted AGN. For two post-merger systems accreting at the Eddington limit, the system that produces a powerful FRII jet is in counterrotation, which makes its gap region between disk and black hole larger. This makes for an initial weaker disk wind and a weaker impact on $\sigma$. FR II sources show an absence of correlation between jet power and stellar dispersion. Hence, jetted AGN tend to begin leftward of non-jetted AGN on the M-$\sigma$ plane and this weaker coupling makes them travel more upwards than horizontally on the diagram. But once the transition to FRI jets is instantiated, such systems generate a dominant effect on $\sigma$. But this tends to occur when accretion rates drop dramatically.  Hence, the most powerful jetted AGN live long periods by moving mostly horizontally in the M-$\sigma$ plane. We have attempted to separate the paths of objects in the M-$\sigma$ plane that react in different ways on star formation, stellar dispersion, accretion, and with different Sersic profiles. Because the FRII to FRI transition in radio galaxies suggests a bending path for radio loud AGN in the M-$\sigma$ plane as shown in Figure 6, we argued that if we go to higher redshift, we will begin to capture more such transition objects and this will result in a bend such as seen in Figure 8. However, if we imagine going to even  higher redshift, to redshift of 2 or so, which often corresponds to the peak in estimated merger functions, one predicts that the high $\sigma$ objects should appear to drop out of the diagram. This is because such objects have not had time to form (i.e. FRI LERG). This will be the subject of future work. Since radio loud AGN are about 10-20 \% of the total AGN population, it is interesting to explore whether the bending we are attributing to them in virial mass versus stellar dispersion (Figure 8) can be achieved by such a small fraction of the population. But if the clustering of the LEGA-C sample is high, the percentage of radio loud AGN would be even higher.

\begin{acknowledgments}
We thank the anonymous referee for interesting insights.
\end{acknowledgments}

\section{References}

Bardeen, J.M.\& Petterson, J.A., 1975, ApJ, 195, L65 

Barisic, I. et al 2017, ApJ, 847, 72

Best, P.N. \& Heckman, T.M., 2012, MNRAS, 421, 1569 

Capetti, A. \& Brienza, M., 2023, arXiv:2306.16078 

Chiaberge, M.\& Marconi, A., 2011, MNRAS, 416, 917

 Fanaroff, B.L. \& Riley, J.M., 1974, MNRAS, 167, 31

Ferrarese, L \& Merritt, D., 2000, ApJ, 539, L9

Garofalo, D., Moravec, E., Macconi, D., Singh, C.B., PASP, 134, 114101

Garofalo, D., Pouliasis, E., 2022, PASP, 134, 094103

Garofalo, D., North, M., Belga, L., Waddell, K., 2020, ApJ, 890, 144

Garofalo, D., Joshi, R., Yang, X., Singh, C.B., North, M., Hopkins, M., 2020, ApJ, 889, 91

Garofalo, D. , Evans, D.A., Sambruna, R.M., 2010, MNRAS, 406, 975

Gebhardt, K. 2000, ApJ, 539, L13

Graham, A.W., Erwin, P., Trujillo, I., \& Asensio Ramos, A. 2003, AJ, 125, 2951

Gültekin, K., Richstone, D. O., Gebhardt, K., et al. 2009, ApJ, 698, 198

Hamilton, T.S., 2010, MNRAS, 407, 2393

Kalfountzou, E. et al 2014, MNRAS, 442, 1181

King, A.R., Lubow, S.H., Ogilvie, G.I., Pringle, J.E., 2005, MNRAS, 363, 49

Kormendy, J. \& Richstone, D.O., 1995, ARA\&A, 33, 581

Macconi, D.. Torresi, E., Grandi, P., Boccardi, B., Vignali, C., 2020, MNRAS, 493, 4355

Magorrian, J. et al 1998, AJ, 115, 2285     

Milosavljevic, M., \& Merritt, D., 2001, ApJ, 563, 34

Mingo, B. et al 2022, MNRAS, 511, 3250 

Mingo, B., Hardcastle, M. J., Croston, J.H., Dicken, D., Evans, D.A., Morganti, R., Tadhunter, C., 2014, MNRAS, 440, 269 

Miraghaei, H. \& Best, P.N., 2017, MNRAS, 466, 4346

Mukherjee, D., Wagner A.Y., Bicknell, G.V., Morganti, R., Oosterloo, T., Nesvadba, N., Sutherland, R.S., 2018, MNRAS, 476, 80

Murphy, J.D., Gebhardt, K., Cradit, M., 2014, ApJ, 785, 143

Neilsen, J. \& Lee, J., Nature, 2009, 458, 481

Nixon, C.J., Cossins, P.J., King, A.R. $\&$ Pringle, J.E., 2011, MNRAS, 412, 1591

Ponti, G. et al 2012, MNRAS, 422, L11

Sahu, N., Graham, A.W. \& David, B.L., 2019, ApJ, 887, 10

Silk, J. \& Rees, M. 1998, A\&A, 331, L1    

Singh, C.B., Kulasiri, N., North, M., Garofalo, D., 2021, PASP, 133, 104101 

Van der Wel, A. et al 2021, ApJS, 256, 44

Zubovas, K \& King, A.R., 2019, GReGr, 51, 65

\end{document}